\documentclass[a4paper,12pt]{article}
\usepackage[dvips]{graphics}
\usepackage[english]{babel}
\usepackage{amssymb,amsmath}
\newcommand{\mC}{\mathbb{C}}

\newcommand{\mR}{\mathbb{R}}
\newcommand{\mT}{\mathbb{T}}
\newcommand{\mS}{\mathbb{S}}
\newcommand{\mZ}{\mathbb{Z}}

\newcommand{\rot}{\text{\rm{rot}}}
\newcommand{\fm}{\phantom{-}}
\newtheorem{theorem}{Theorem}
\newtheorem{definition}[theorem]{Definition}
\newtheorem{proposition}[theorem]{Proposition}
\newtheorem{lemma}[theorem]{Lemma}
\newtheorem{corollary}[theorem]{Corollary}

\newtheorem{remark}[theorem]{Remark}

\newcommand{\qed}{$\Box$}

\title{Cantor Spectrum for the Almost Mathieu Operator.
 Corollaries of localization, reducibility and duality}
\author{Joaquim Puig${}^{(1)}$}
\date{}
\begin{document}

\maketitle
{\footnotesize{\hspace*{4mm}(1) Dept. de Matem\`atica Aplicada i
An\`alisi, Univ. de Barcelona, Gran Via 585, 08007 Barcelona,\\
\hspace*{15mm} Spain}}

\maketitle
\begin{abstract}
In this paper we use results on reducibility, localization and duality
for the Almost Ma\-thieu operator,
$$ \left(H_{b,\phi} x\right)_n= x_{n+1} +x_{n-1} + b \cos\left(2 \pi n \omega +
\phi\right)x_n $$ 
on $l^2(\mathbb{Z})$ and its associated eigenvalue equation to deduce 
that for $b \ne 0,\pm 2$ and
$\omega$ Diophantine the spectrum of the operator is a Cantor subset of the
real line. This solves the so-called ``Ten Martini Problem''
for these values of $b$ and $\omega$. Moreover, we prove that 
for $|b|\ne 0$ small enough or large enough 
all spectral gaps predicted by the Gap Labelling theorem are open.
\end{abstract}
\section{Introduction. Main results}
\label{intro}

In this paper we study the nature of the spectrum of the Almost Mathieu operator
\begin{equation}\label{eq:almostop}
(H_{b,\phi} x)_n = x_{n+1}+x_{n-1} + b\cos(2\pi \omega n + \phi) x_n, \qquad n \in \mathbb{Z}
\end{equation}
on $l^2(\mathbb{Z})$, where $b$ is a real parameter, $\omega$ is an irrational number and
$\phi\in \mathbb{T}= \mathbb{R}/(2\pi\mathbb{Z})$. Since for each $b$ this is a bounded self-adjoint operator, the spectrum
is a compact subset of the real line which does not depend on $\phi$ because of the
assumption on $\omega$. This spectrum will be denoted by $\sigma_b$.
For $b=0$, this set is the interval $[-2,2]$. The understanding of the
spectrum of (\ref{eq:almostop}) is related to the dynamical
properties of the difference
equation
\begin{equation}\label{eq:almosteq}
 x_{n+1}+x_{n-1} + b\cos(2\pi \omega n + \phi) x_n= a x_n, \qquad n \in \mathbb{Z}
\end{equation}
for $a \in \mathbb{R}$, which is sometimes called the Harper equation. In what follows we will assume that the frequency $\omega$ is Diophantine:

\begin{definition}
We say that a real number $\omega$ is Diophantine whenever there exist
positive constants $c$ and $r>1$ such that the estimate
$$ \left| \sin{2\pi n  \omega}\right| > \frac{c}{|n|^r} $$
holds for all $n\ne 0$.
\end{definition}

The nature of the spectrum of this operator has been studied intensively
in the last twenty years (for a review, see Last \cite{last:almost}) and
an open problem has been to know whether the spectrum is a Cantor
set or not, which is usually referred as the ``Ten Martini
Problem''. In this  paper we derive two results  on this problem. 
The first one is \emph{non-perturbative}:

\begin{corollary}\label{res:debil}
If $\omega$ is Diophantine, then the spectrum of the Almost Mathieu 
operator is a Cantor set if $b \ne 0,\pm 2$.
\end{corollary}

Here, we prefer to call this result a corollary, rather than a theorem,
because the proof requires just a combination of reducibility, point
spectrum and duality developed quite recently for the Almost Mathieu
operator and the related eigenvalue equation. The argument
 is in fact reminiscent of Ince's original argument for the
classical Mathieu differential equation (see \cite{ince}). 
In the critical case $|b| = 2$,  Y. Last proved in \cite{last:zero} 
that the spectrum of the
Almost Mathieu operator is a subset of the real line with zero Lebesgue
measure and that it is a Cantor set for the values of $\omega$ which
have an unbounded continued fraction expansion, which is a set of full
measure. This last result has been extended recently to all
Diophantine frequencies by Avila \& Krikorian \cite{avila-krikorian}.

The Cantor structure of the spectrum of the Almost Mathieu operator can
be better understood if we make use of the concept of \emph{rotation
number}, which can be defined as follows. Let $(x_n)_{n\in
\mathbb{Z}}$ be a non-trivial solution of (\ref{eq:almosteq}),
for some fixed $a,b,\phi.$ Let $S(N)$ be the number of changes
of sign of such solution for $1 \le n \le N$, adding one if
$x(N)=0$. Then the limit 
\[
\lim_{N \to \infty} \frac{S(N)}{2 N}
\]
exists, it does not depend on the chosen solution $x$, nor on
$\phi$ and it is denoted by $\rot(a,b)$. A more complete
presentation of this object can be found in section
\ref{sec:prerequisites}. Here we only mention some
properties which relate it to the spectrum of $H_{b,\phi}$: 

\begin{proposition}[\cite{avron-simon,delyon-souillard,herman,johnson-moser}]\label{res:propietats}
The rotation number has the following properties:
\begin{itemize}
\item[(i)] The rotation number, $\rot(a,b)$, is a continuous function of $(a,b) \in
\mathbb{R}^2.$
\item[(ii)] For a fixed $b$, the spectrum of
$(\ref{eq:almostop})$, $\sigma_b$, is the set of $a_0\in
\mathbb{R}$, such that $a \mapsto \rot(a,b)$ is not locally
constant at $a_0$.
\item[(iii)] $(${\bf{Gap labelling}}$)$ If $I$ is an open,
non-void interval in the resolvent set of
$(\ref{eq:almostop})$, $\rho_b= \mathbb{R}-\sigma_b$, then
there is an integer $k \in \mathbb{Z}$ such that 
\[
2 \rot(a,b) - k \omega \in \mathbb{Z} 
\]for all $a \in I.$
That is,
\[
\rot(a,b)=\frac{1}{2}\left\{ {k\omega} \right\}
\] where $\{\cdot\}$ denotes the fractional part of a
real number.
\end{itemize}
\end{proposition}

\begin{figure}
\begin{center}
{\resizebox{10.5cm}{!}{\includegraphics{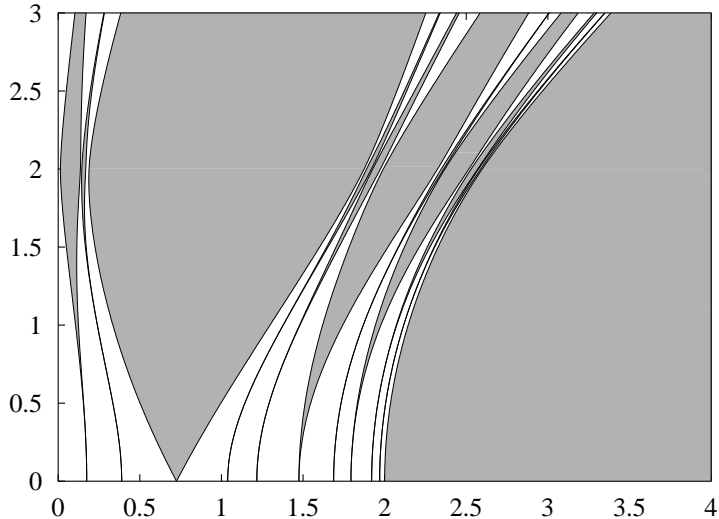}}}
\vspace{-2mm}
{\caption{\small{Numerical computation of the  ten biggest spectral 
gaps for the Almost Mathieu operator with different values of $b$ and
$\omega= (\sqrt{5}-1)/2$. They correspond to the first  $|k|$
such that $\{k \omega\}/2$ belongs to $[1/4,1/2]$.
The coupling parameter $b$ is in the vertical direction whereas the
spectral one, $a$, is in the horizontal one. Note that for $b=0$, all gaps 
 except the upper one are collapsed.}}
\label{fig:dualitats}}
\end{center}
\vspace*{-2mm}
\end{figure}

From this theorem we conclude that the resolvent set is the disjoint union of 
countably (or finitely) many open
intervals called \emph{spectral gaps}, possibly void, and which can be
uniquely labelled by an integer $k$ called the
\emph{resonance}. If the closure of a spectral gap
degenerates to a point we will say that it is a
\emph{collapsed gap} and
otherwise that it is a \emph{non-collapsed gap}. See Figure
\ref{fig:dualitats} for a numerical computation of the
biggest gaps in  the spectrum of the Almost Mathieu operator 
for several values $b$.

In particular if, for a fixed $b$, all the spectral gaps are
open and the frequency $\omega$
is irrational, then the spectrum $\sigma_b$ is a Cantor set. 
The question of the non-collapsing
of all spectral gaps is sometimes called the \emph{Strong
(or Dry) Ten Martini Problem}. However, if non-collapsed
gaps are dense in the spectrum, then  this is still a Cantor set, 
although some (perhaps an infinite number) of collapsed gaps may also coexist. 

Now we can formulate the second corollary in this paper:

\begin{corollary}\label{res:fort}
Assume that $\omega \in \mathbb{R}$ is Diophantine. Then, there 
is a constant $C=C(\omega)>0$ such
that if $0<|b|<C$ or $4/C < |b| <\infty$ all the spectral 
gaps of the spectrum of the Almost Mathieu operator are open.
\end{corollary}

Before ending this introduction we  give a short account of  the existing
results (to our knowledge) on the Cantor spectrum of 
the Almost Mathieu operator 
for $|b|\ne 0,2$. The Cantor spectrum for the Almost Mathieu operator
was first conjectured by Azbel \cite{azbel} and Kac, in 1981,
conjectured that all the spectral gaps are open. The problem of the Cantor 
structure of the spectrum was called the ``Ten Martini Problem'' by Simon
\cite{simon:review} (and remained as Problem 4 in \cite{simon:sXXI}). 
Sinai \cite{sinai}, proved that for
Diophantine $\omega$'s and sufficiently large (or small $|b|$),
depending on $\omega$, the spectrum $\sigma_b$ is a Cantor set. Choi,
Elliott \& Yui \cite{choi-elliott-yui} proved that the spectrum
$\sigma_b$ is a Cantor set for all $b \ne 0$ when $\omega$ is a Liouville
number obeying the condition
$$ \left| \omega - \frac{p}{q} \right| < D^{-q},$$
for a certain constant $D>1$ and infinitely many rationals $p/q$. In
particular, this means that for a $G_\delta$-dense subset of pairs
$(b,\omega)$ the spectrum is a Cantor set, which is the Bellissard-Simon
result \cite{bellissard-simon}. For generic results on
Cantor spectrum for almost periodic an quasi-periodic
Schr\"odinger operators see Moser \cite{moser} and Johnson
\cite{johnson-91}. Nevertheless, collapsed gaps appear
naturally in quasi-periodic Schr\"odinger operators, as it was
shown by Broer, Puig \& Sim\'o \cite{broer-puig-simo} and
there are examples which do not display Cantor spectrum, see
De Concini \& Johnson \cite{deconcini-johnson}. Finally, let us 
mention that, if we 
consider the case of rational $\omega$, all spectral gaps, apart from the middle one, are open if $b\ne  0$. This result was proved by van Mouche \cite{mouche}
and Choi, Elliott \& Yui \cite{choi-elliott-yui}.

Let us now  outline the contents of the present paper. In Section
\ref{sec:prerequisites} we introduce some of the tools
needed to prove our two main results. These include the
different definitions of the rotation number, the concept of
reducibility of linear quasi-periodic skew-products and
the duality for the Almost Mathieu operator. In Section
\ref{sec:fort} apply the reducibility results by Eliasson to
prove Corollary \ref{res:fort}. Finally, in Section
\ref{sec:debil}, the proof of of Corollary \ref{res:debil}
is given, which is based on a result of non-perturbative
localization by Jitomirskaya.

\section{Prerequisites: rotation number,
reducibility, duality and lack of coexistence}\label{sec:prerequisites}

%%%Rotation number
\subsection*{Rotation number}
The rotation number for quasi-periodic Schr\"odinger
equations is a very useful object with  deep connections to
the spectral properties of Schr\"odinger operators. It is
also related to the dynamical properties of the solutions of
the associated eigenvalue equation. This
allows several equivalent definitions, which we shall now
try to present.

The rotation number was introduced for continuous time
quasi-periodic Schr\"o\-din\-ger equations by Johnson \& Moser
\cite{johnson-moser}. The discrete version was introduced by
Herman \cite{herman} (which is also defined for quasi-periodic 
skew-product flows on $SL(2,\mR) \times \mT$) and Delyon \& Souillard
\cite{delyon-souillard} (which is the definition given in
the introduction). We will now review these definitions, 
their connection and some important properties.

Herman's definition is dynamical. Here we follow the
presentation by Krikorian \cite{krikorian:sl2}. Write 
equation (\ref{eq:almosteq}) 
as a \emph{quasi-periodic skew-product flow} on 
$\mathbb{R}^2 \times \mathbb{T},$
\begin{equation}\label{eq:skewgeneral}
u_{n+1}= A(\theta_n) u_n
\qquad \theta_{n+1} = \theta_n + 2\pi
\omega,
\end{equation}
setting $u_{n}= \left(x_{n},x_{n-1}\right)^T$ and 
\begin{equation}\label{eq:defa}
A(\theta)= \left(
\begin{array}{cc}
a -b \cos{\theta}  & \; -1 \\
1 & \; \fm 0
\end{array}
\right),
\end{equation}
which belongs to $SL(2,\mathbb{R})$ the group of bidimensional
matrices with determinant one. The quasi-periodic flow can
also be defined on $SL(2,\mR) \times \mT$ considering the
flow given by
\begin{equation}\label{eq:skewlift}
X_{n+1}= A(\theta_n) X_{n},\qquad \theta_{n+1} = \theta_n +
2\pi \omega,
\end{equation}
with $X_0 \in SL(2,\mR)$. This can be seen as the
iteration of the following \emph{quasi-periodic cocycle} on
$SL(2,\mR) \times \mT$:
\begin{equation}\label{eq:cocycle}
\begin{array}{rcl}
SL(2,\mR) \times \mT & \longrightarrow & SL(2,\mR) \times \mT \\
 \left(X, \theta \right) & \mapsto & \left(A(\theta) X,
 \theta + 2\pi \omega\right),
\end{array}
\end{equation}
which we denote by $(A,\omega)$. 

We will now give Herman's definition of the rotation number
of a quasi-periodic cocyle like (\ref{eq:cocycle}) with
$A:\mT \to SL(2,\mR)$ homotopic to the identity. For a
general $A:\mT \to SL(2,\mR)$, this last property is
not always true, since $SL(2,\mR)$ is not simply
connected. Indeed, its first homotopy group is isomorphic to
$\mZ$, with
generator the rotation $R_1: \mT \to SL(2,\mR)$ given by
\[
R_1(\theta)=\left(
\begin{array}{cc}
\cos{\theta} & -\sin{\theta} \\
\sin{\theta} & \phantom{-}\cos{\theta}
\end{array} \right)
\]
for all $\theta \in \mT$. In our case, the 
\emph{Almost Mathieu cocyle} (\ref{eq:defa}) is homotopic to
the identity. 

Let $\mS^1$ be the set of unit vectors of $\mR^2$ and let us
denote by $p:\mR \to \mS^1$ the projection given by the
exponential $p(t)= e^{i t}$, identifying $\mR^2$ with $\mC$.
Because of the linear character of the cocyle, the continuous map
\begin{equation}
\begin{array}{rrcl}
F: & \mS^1 \times \mT  & \longrightarrow     & \mS^1 \times \mT \\
      & (v,\theta)    & \mapsto & \left(
      \frac{A(\theta)v}{\| A(\theta) v \|}, \theta+2\pi\omega \right)
\end{array}
\end{equation}
is also homotopic to the identity. Therefore, it admits a continuous lift
$\tilde{F}: \mR \times \mT \to  \mR \times \mT$ of the form:
\[
\tilde{F}(t,\theta)= \left( t+ f(\theta,t), \theta + 2\pi\omega\right)
\]
such that
\[
f(t +2\pi,\theta + 2\pi\omega)= f(t,\theta) \text{ and } p\left(t + f(t,\theta)\right)=
\frac{A(\theta)p(t)}{\| A(\theta) p(t) \|}
\]
for all $t \in \mR$ and $\theta \in \mT$. The map $f$ is independent of the
choice of $\tilde{F}$ up to the addition of a constant $2\pi k$, with $k \in
\mZ$. Since the map $\theta \mapsto \theta + 2\pi\omega$ is
uniquely ergodic on $\mT$ for all $(t,\theta) \in
\mR\times\mT$, the limit 
\[
\lim_{N \to \infty} \frac{1}{2\pi N} \sum_{n=0}^{N-1}
f\left(\tilde{F}^n(t,\theta)\right)
\]
exists, it is independent of $(t,\theta)$ and the convergence
is uniform in $(t,\theta)$, see Herman \cite{herman} and Johnson \& Moser
\cite{johnson-moser}. This object is called the
\emph{fibered rotation number}, which will be denoted as $\rho_f(a,b)$, 
and it is defined modulus $\mZ$.

For instance, if $A_0 \in SL(2,\mR)$
is a constant matrix, then the
rotation number of the cocycle $(A_0,\omega)$, for any irrational
$\omega$, is the absolute value of the argument of the
eigenvalues divided by $2\pi$.  

Using a suspension argument (see Johnson \cite{johnson:review}) 
it can be seen that, for the Almost Mathieu cocycle (like for 
any quasi-periodic Schr\"odinger cocycle), the fibered
rotation number coincides with the \emph{Sturmian} definition given in the
introduction. Note that this last rotation number,
${\text{rot}}(a,b)$, belongs to the interval, whereas the
fibered rotation number, $\varrho_f(a,b)$, is an element of
$\mR/\mZ$. They can be both linked by means of the \emph{integrated
density of states}, see Avron \& Simon \cite{avron-simon}. 
Let $k_L(a,b,\phi)$ be $(L-1)^{-1}$ times the number of eigenvalues
less than or equal to $a$ for the restriction of
$H_{b,\phi}$ to the set $\{1,\ldots,L-1\}$, for some $\phi
\in \mT$, with zero boundary conditions at both ends $0$ and
$L$. Then, as $L \to \infty$, the $k_{L}(a,b,\phi)$ converge to a
continuous function $k(a,b)$, which is the integrated density
of states. The basic relations are
\[
2 \rot(a,b)= k(a,b) \qquad{\text{and}} \qquad 2\varrho_f(a,b)=
k(a,b) + l,
\]
for a suitable integer $l \in \mZ$. In particular, 
\[
\rot(a,b)= \rho_f(a,b) \text{ (mod } \frac{1}{2}\mZ ).
\]

In what follows, the arithmetic nature of the rotation
number will be of importance. We will say that the rotation
number is \emph{rational or resonant} with respect to
$\omega$ if there exists a constant $k \in \mZ$ such that
$\rot(a,b)= \{k\omega\}/2$ or equivalently, $\varrho_f(a,b)=
k\omega/2$ modulus $\frac{1}{2}\mZ$. Also, we say that it is
\emph{Diophantine} with respect to $\omega$ whenever the
bound
\[
\left|\rot(a,b)-\frac{\{k\omega\}}{2}\right|= \min_{l \in
\mZ}\left|\rho_f(a,b)-\frac{k \omega}{2}-\frac{l}{2}\right|
\ge \frac{K}{|k|^\tau},
\]
holds for all $k \in \mZ-\{0\}$ and suitable fixed positive
constants $K$ and $\tau$.

\subsection*{Reducibility}

A main tool in the study of quasi-periodic skew-product
flows is its \emph{reducibility} to constant coefficients. 
Reducibility  is a concept defined for the continuous and discrete
case (for an introduction see the reviews by Eliasson
\cite{eliasson:budapest}, \cite{eliasson:berlin} and, for
more references, the survey \cite{puig:reducibility} by the author).

A quasi-periodic skew-product 
flow like (\ref{eq:skewgeneral}), or a quasi-periodic
cocycle like (\ref{eq:cocycle}), with $A:\mathbb{T}\to
SL(2,\mR)$, is said to be \emph{reducible to 
constant coefficients} if there is a continuous map
$Z:\mathbb{T} \to SL(2,\mathbb{R})$ and a constant matrix
$B \in SL(2,\mathbb{R})$, called the
Floquet matrix, such that the conjugation
\begin{equation}\label{eq:conjugation}
A(\theta) Z(\theta) = Z(\theta + 2\pi \omega) B
\end{equation}
is satisfied for all $\theta \in \mathbb{T}.$ 
When $\omega$ is rational, in which
case the flow is periodic, any skew-product flow is reducible
to constant coefficients. Even in this periodic case, it is not always possible 
to reduce with the same frequency $\omega$, but with $\omega/2$.
If there is a reduction to constant coefficients like
(\ref{eq:conjugation}), then a fundamental matrix of solutions
of (\ref{eq:skewgeneral}), 
\[
X_{n+1}(\phi)= A(2\pi n\omega+\phi)X_n(\phi),\qquad n\in \mathbb{Z},
\]
with $X_0:\mT \to SL(2,\mathbb{R})$ continuous, has the following 
\emph{Floquet representation}:
\begin{equation}\label{eq:floquetrepresentation}
X_n(\phi)= Z(2\pi n \omega + \phi) B^n Z(\phi)^{-1} X_0(\phi)
\end{equation}
for all $n\in  \mathbb{Z}$ and $\phi \in \mathbb{T}$. This
gives a complete description of the qualitative behaviour of
the flow (\ref{eq:skewgeneral}). 

The rotation number of a quasi-periodic cocycle is not
invariant through a conjugation like (\ref{eq:conjugation}).
There are however the following easy relations:

\begin{proposition}
Let $\omega$ be an irrational number and $(A_1,\omega)$ and
$(A_2,\omega)$ be  two quasi-periodic cocycles
on $SL(2,\mR) \times $ homotopic to the identity, being
$\rho_1$ and $\rho_2$ the corresponding fibered 
rotation numbers. Assume that there
exists a continuous map $Z: \mT \to SL(2,\mR)$ such that
\[
A_1(\theta) Z(\theta) = Z(\theta+ 2\pi\omega) A_2(\theta)
\]
for all $\theta \in \mT$. Then, if $k \in \mZ$ is the degree
of $Z$,
\[
\rho_1= \rho_2 + k \alpha \text{ modulus } \mZ.
\]
\end{proposition}

This proposition shows that, for any fixed irrational
frequency $\omega$, the class of quasi-periodic cocycles with
rational rotation number (resp. with Diophantine rotation
number) is invariant under conjugation, although the
rotation number itself may change. Also,
that whenever a quasi-periodic skew-product flow 
 in $SL(2,\mR) \times \mT$ is reducible to a
Floquet matrix with trace $\pm 2$, the rotation number
must be rational.

\subsection*{Duality and lack of coexistence}

To end this section, let us present a specific feature of the Almost
Mathieu operator or, rather, of the associated eigenvalue
equation which will in the basis of our arguments. 
It is part of what is known as \emph{Aubry duality} or simply \emph{duality}:

\begin{theorem}[Avron \& Simon \cite{avron-simon}]\label{res:dualrot}
For every irrational $\omega$, the rotation number of (\ref{eq:almosteq}) satisfies the
relation
\begin{equation}\label{eq:dualitat}
\rot(a,b)= \rot(2a/b,4/b) 
\end{equation}
for all $b \ne 0$ and $a \in \mathbb{R}$.
\end{theorem}

According to Proposition \ref{res:propietats} this means that 
the spectrum $\sigma_{4/b}$, for $b\ne 0$ is just a
dilatation of the spectrum $\sigma_b$. In particular, $\sigma_b$ 
is a Cantor set (resp. none of
the spectral gaps of $\sigma_b$ is collapsed) if, and only if
$\sigma_{4/b}$ is a Cantor set (resp. none of the spectral gaps of
$\sigma_{4/b}$ is collapsed).

In the proof of our two main results we will use the
following argument, which is analogous to Ince's argument
for the classical Mathieu periodic diffe\-rential equation
(see \cite{ince} \S 7.41). In principle, the eigenvalue equation of a
general  quasi-periodic Schr\"odinger operator may have
 two linearly independent quasi-periodic solutions with frequency
$\omega$ (or $\omega/2$). One may call this phenomenon
\emph{coexistence} of quasi-periodic solutions, in analogy
with the classical Floquet theory for second-order periodic
differential equations. A trivial example of this occurs in the
Almost Mathieu case for $b=0$ and suitable values of $a$.

Let us now show that in the Almost Mathieu case this does
not happen if $b \ne 0$, i.e. two
quasi-periodic solutions with frequency $\omega$ of the
eigenvalue equation cannot coexist. Let $(x_n)_{n\in \mZ}$ 
satisfy the equation
\begin{equation}\label{eq:almostintro}
x_{n+1} + x_{n-1} + b \cos(2\pi\omega n + \phi) x_n = a x_n,
\qquad n \in \mZ
\end{equation}
for some $a$, $b\ne 0$ and $\phi$. If it is quasi-periodic with
frequency $\omega$, there exists a continuous function $\psi: \mT \to
\mR$ such that $x_n= \psi(2\pi\omega n + \phi)$ for all $n \in
\mZ$. The Fourier coefficients of $\psi$, $(\psi_m)_{m\in
\mZ}$ satisfy the following equation:
\[
2 \cos(2\pi\omega m) \psi_m + \frac{b}{2} (\psi_{m+1} +
\psi_{m-1})
= a \psi_m,\qquad m\in \mathbb{Z},
\]
which is equivalent to 
\begin{equation}\label{eq:almosteqdual}
 \psi_{m+1} + \psi_{m-1} + \frac{4}{b} \cos(2\pi\omega m)
 \psi_m   = \frac{2a}{b} \psi_m, \qquad m\in \mathbb{Z}.
\end{equation}

Since $\psi$ is at least continuous, then $(\psi_m)_{m\in\mZ}$
belongs to $l^2(\mZ)$. Now the reason for the absence of
coexisting quasi-periodic solutions is clear. Indeed, if 
if $(y_n)_{n\in\mZ}$ is another linearly independent
quasi-periodic solution  of (\ref{eq:almostintro}) with frequency $\omega$, say
$y_n=\chi(2\pi\omega + \phi)$, for some continuous $\chi$,
then the sequence of the Fourier coefficients of $\chi$,
$(\chi_{m})_{m\in\mZ}$, would be a solution of
(\ref{eq:almosteqdual}) belonging to $l^2(\mZ)$. 
The sequences $(\psi_m)_{m\in\mZ}$ and $(\chi_m)_{m\in\mZ}$
would be two linearly independent solutions of
(\ref{eq:almosteqdual}) which belong both to $l^2(\mZ)$.

This is a contradiction, because for bounded potentials, like the cosine, 
we are always in the limit-point case 
(see \cite{carmona-lacroix,coddington-levinson} for the
continuous case). In our discrete case, this is even simpler, since
any solution in $l^2(\mathbb{Z})$ of the eigenvalue equation must tend to
zero at $\pm \infty$. Hence, the existence of two linearly
independent solutions belonging both to $l^2(\mathbb{Z})$ would be in
contradiction with the preservation of the Wronskian. 

Therefore, two quasi-periodic solutions with frequency
$\omega$ cannot coexist if $b\ne 0$. A similar argument
shows that quasi-periodic solutions of the form
\begin{equation}\label{eq:antiperiodic}
(-1)^n\psi(2\pi\omega n + \phi),
\end{equation}
for a continuous $\psi:\mT \to \mR$ cannot coexist.

Finally, note that the coexistence of two quasi-periodic
solutions with frequency $\omega$ of equation
(\ref{eq:almostintro}) is equivalent to the reducibility of
the corresponding two-dimensional skew-product flow
(\ref{eq:skewgeneral}), with the identity as Floquet matrix. 
Similarly the coexistence of two quasi-periodic solutions of the type
(\ref{eq:antiperiodic}) is equivalent to the reducibility of
the flow with minus the identity as Floquet matrix.

\section{The Strong Ten Martini Problem for small (and
large) $|b|$}\label{sec:fort}

In this section we will show that for $0<|b|<C$, where $C>0$ is a
suitable constant, and for $|b|>4/C$ all  spectral gaps are open. 

The theorem from which we will derive
Corollary \ref{res:fort} is due to Eliasson and it was
originally stated for the continuous case, based on a KAM
scheme. It can be adapted to the discrete case to obtain
the following:

\begin{theorem}[\cite{eliasson:sagaro,eliasson:floquet}]\label{res:eliasson}
Assume that $\omega$ is Diophantine with constants $c$ and $r$. 
Then there is a constant $C(c,r)$ such that, if $|b|<C(c,r)$ 
and $\rot(a,b)$ is either rational or Diophantine,
then the quasi-periodic skew-product flow 
\begin{equation}\label{eq:skew}
\left(\begin{array}{c}
x_{n+1} \\
x_{n} 
\end{array}\right)=
\left(
\begin{array}{cc}
a -b \cos{\theta_n}  & \; -1 \\
1 & \; \fm 0
\end{array}
\right)
\left(\begin{array}{c}
x_{n} \\
x_{n-1} 
\end{array}\right),\qquad \theta_{n+1} = \theta_n + 2\pi \omega
\end{equation}
on $\mathbb{R}^2 \times \mathbb{T}$ is reducible to constant 
coefficients, with Floquet matrix $B$, by means of
a quasi-periodic (with frequency $\omega/2$)
and analytic transformation. Moreover, if $a$ is at an endpoint of a
spectral gap of $\sigma_b$, then the trace of $B$ is $\pm 2$, 
being $B=\pm I$ if, and only if, the gap collapses. Finally,
if $B= \pm I$ then the transformation $Z$  can be chosen to 
have frequency $\omega$.
\end{theorem}

For other reducibility results in the context of
quasi-periodic Schr\"odinger operators see Dinaburg \& Sinai
\cite{dinaburg-sinai} and Moser \& P\"oschel
\cite{moser-poschel} for the continuous case and Krikorian
\cite{krikorian:sl2} and Avila \& Krikorian
\cite{avila-krikorian} for the discrete case.

Taking into account the arguments from the previous section,
Corollary \ref{res:fort} is immediate. Indeed, let 
$|b|<C$, where $C$ is the constant given by the theorem
for a fixed Diophantine frequency $\omega$.
Then the skew-product flow (\ref{eq:skew}) is reducible to constant
coefficients and the Floquet matrix has trace $\pm 2$ if $a$
is an endpoint of a spectral gap. Moreover the gap is
collapsed if, and only if, the Floquet matrix $B$ is 
$\pm I$. Since we
have seen in the previous section that (\ref{eq:skew}) for
$b\ne0$ cannot be reducible to these Floquet matrices,
Corollary \ref{res:fort} follows.

\section{Non-perturbative localization and Cantor spectrum for
$b\ne0$}\label{sec:debil}

In this section we will see how Corollary \ref{res:debil} is
a consequence of the following theorem on non-perturbative
localization, due to Jitomirskaya:

\begin{theorem}[\cite{jito:metal}]\label{res:jito}
Let $\omega$ be Diophantine. Define the set $\Phi$ of \emph{resonant phases} as the set of those $\phi \in \mathbb{T}$ such that the
relation
\begin{equation}\label{eq:resonant}
\left| \sin \left( \phi + \pi n \omega  \right) \right| < \exp \left( - |n|^{\frac{1}{2r}}
\right) 
\end{equation}
holds for infinitely many values of $n$, being $r$ the constant in the
definition of a Diophantine number. Then, if $\phi \not\in \Phi$ and $|b|>2$ the
operator $H_{b,\phi}$ has only pure point spectrum with exponentially decaying
eigenfunctions. Moreover, any of these eigenfunctions $(\psi_n)_{n\in \mathbb{Z}}$ satisfies that
\begin{equation}\label{eq:ressonancia}
\beta(b)= -\lim_{|n| \to \infty} \frac{\log\left(\psi^2_n + \psi^2_{n+1}\right) }{2|n|} =
\log\left(\frac{|b|}{2}\right). 
\end{equation}
\end{theorem}

Now we prove Corollary \ref{res:debil}. Let $|b|>2$. Then, 
according to Theorem \ref{res:jito}, 
the operators $H_{b,0}$ and $H_{b,\pi/2}$ have only pure point spectrum 
with exponentially decaying eigenfunctions. The
eigenvalue equation associated to these operators has the following properties:

\begin{lemma}\label{res:sym}
Let $(x_n)_{n \in \mathbb{Z}}$ be a solution of the difference equation
$$  x_{n+1} +x_{n-1} + b\cos(2\pi n \omega+ \phi) x_n = a x_n, \qquad n \in \mathbb{Z},$$
for some constants $a,b$ and $\phi \in \mathbb{T}$. Then, if $\phi=0$, $(x_{-n})_{n \in \mathbb{Z}}$ is also a solution of this equation and, if $\phi= -\pi/2$ then $(-x_{-n})_{n \in
\mathbb{Z}}$ is also a solution.
\end{lemma}

Let us
consider first the case of the operator $H_{b,0}$.  According to
Theorem \ref{res:jito}, 
there exists a sequence of eigenvalues $(a^k(b))_{k \in \mathbb{Z}}$ with
eigenvectors $(\psi^k(b))_{k \in \mathbb{Z}} $, exponentially localized and
which form a complete orthonormal basis of $l^2(\mathbb{Z})$. Moreover the set of 
eigenvalues $(a^k(b))_{k \in \mathbb{Z}}$ must  be dense in the spectrum $\sigma_b$. 
Again, we do not write the dependence on $b$ for simplicity in what
follows. None of these eigenvalues can be repeated, since we are in the limit
point case. Writing each of the $\psi^k$ as
\[
\psi^k = (\psi^k_n)_{n \in \mathbb{Z}},
\]
we define
\[
{\tilde \psi}^k(\theta)= \sum_{k \in \mathbb{Z}} \psi^k_n e^{i k
\theta},
\]
for $\theta \in \mathbb{T}$. All these functions belong to
$C^a_{\beta}(\mathbb{T},\mathbb{R})$, the set of real analytic functions of
$\mathbb{T}$ with analytic extension to $|\Im \theta|<\beta$ and they are even
functions of $\theta$, because of Lemma \ref{res:sym} (here we have applied 
again that we are in the limit point case). Passing to the dual equation, 
we obtain that, for each $k \in \mathbb{Z}$,
the sequence $({\tilde \psi}^k( 2\pi \omega n))_{n \in \mathbb{Z}}$ is a
quasi-periodic solution of
\begin{equation}\label{eq:partdual}
 x_{n+1} + x_{n-1} + \frac{4}{b} \cos{\theta_n} x_n = \frac{2 a}{b}x_n,
\qquad \theta_{n+1} =
\theta_n + 2\pi \omega \qquad n \in \mathbb{Z},
\end{equation}
provided $a$ is now replaced by $a^k$. We are now going to see that
$2{a^k}/b$ is at an endpoint of a spectral gap and that this is
collapsed.  To do so we will use reducibility as in the proof
of Theorem \ref{res:fort}. For a direct proof that $2{a^k}/b$ is at
an endpoint of a gap (it has rational rotation number), see again Herman \cite{herman}.

The fact that $({\tilde \psi}^k( 2\pi \omega n))_{n \in \mathbb{Z}}$
is a quasi-periodic solution of (\ref{eq:partdual}) means that, for
all $\theta \in \mT^d$, the following equation is satisfied
\[
\left(
\begin{array}{c}
{\tilde \psi}^k( 4\pi \omega  + \theta) \\
{\tilde \psi}^k( 2\pi \omega  + \theta)
\end{array}
\right)=\left( 
\begin{array}{cc}
\frac{2 a}{b} - \frac{4}{b} \cos{\theta} & \; -1 \\
1                                        & \; \fm 0 
\end{array}\right)
\left(\begin{array}{c}
{\tilde \psi}^k( 2\pi \omega  + \theta) \\
{\tilde \psi}^k(  \theta)
\end{array}
\right).
\]
The following lemma shows that, if this is the case, then the
quasi-periodic skew-product flow 
\begin{equation}\label{eq:skewdebil}
\left(\begin{array}{c}
x_{n+1} \\
x_{n} 
\end{array}\right)=
\left(
\begin{array}{cc}
\frac{2 a^k}{b} - \frac{4}{b} \cos{\theta_n}  & \;  -1 \\
1 & \; \fm 0
\end{array}
\right)
\left(\begin{array}{c}
x_{n} \\
x_{n-1} 
\end{array}\right),\qquad \theta_{n+1} = \theta_n + 2\pi \omega
\end{equation}
is reducible to constant coefficients.

\begin{lemma}
Let $A : \mathbb{T} \to SL(2,\mathbb{R})$ be a real analytic map, with analytic
extension to $|\Im \theta| < \delta$ for some $\delta>0$.
Assume that there is a nonzero real analytic map $v: \mathbb{T} \to \mathbb{R}^2$, 
with analytic extension to $|\Im \theta| < \delta$ such that 
\[
v(\theta + 2\pi \omega) = A(\theta) v(\theta)
\]
holds for all $\theta \in \mathbb{T}.$ Then, the quasi-periodic skew-product 
flow given by
\begin{equation}\label{eq:skewlemma}
u_{n+1}= A(\theta_n) u_n,\qquad \theta_{n+1}= \theta_n + 2\pi\omega,
\end{equation}
with $(u_n,\theta_n) \in \mR^2\times\mT$ for all $n\in \mZ$
is reducible to constant coefficients by means of a quasi-periodic transformation which
is analytic in $|\Im \theta| < \delta$ and has frequency $\omega$. Moreover the 
Floquet matrix can be chosen to be of the form
\begin{equation}\label{eq:floquetform}
B= \left(
\begin{array}{cc}
1 & \; c \\
0 & \; 1 
\end{array}
\right)
\end{equation}
for some $c \in \mR$.
\end{lemma}

\noindent{\bf{Proof:}}
Since $v=(v_1,v_2)^T$ does not vanish, $d=v_1^2+v_2^2$ is
always different from zero and the transformation
\[
Z(\theta)=\left(\begin{array}{cc}
v_1(\theta) & \; -v_2(\theta)/d(\theta) \\
v_2(\theta) & \; \fm v_1(\theta)/d(\theta) 
\end{array}\right),
\]
is an analytic map $Z:\mT \to SL(2,\mR)$. The  transformation 
$Z$ defines a conjugation of $A$ with $B^1$, being
\[
 A(\theta) Z(\theta)= Z(2\pi\omega+\theta)B^1(\theta) , 
\]
which means that $B^1$ is
\[ 
B^1(\theta)= 
\left(\begin{array}{cc}
1 & \; b^1_{12} (\theta) \\
0 & \; 1  
\end{array}\right),
\]
for some analytic $b^1_{12}:\mT \to \mR$. The transformed
skew-product flow, defined by the matrix $B^1$ is reducible
to constant coefficients because it is in triangular form, 
the frequency $\omega$ is Diophantine and $b^1_{12}$ is
analytic. Indeed, if $y_{12}:\mT \to \mR$ is an analytic solution
of the small divisors equation
\[
y_{12}(2\pi\omega +\theta) - y_{12}(\theta) =
b^1_{12}(\theta)-[b^1_{12}],\qquad \theta \in \mT,
\]
where $[b^1_{12}]$ is the average of $b^1_{12}$ (see
\cite{arnold:geometrical}), then the
transformation
\[
Y(\theta)=\left(\begin{array}{cc}
1 & \; y_{12} \\
0 & \; 1  
\end{array}\right)
\]
conjugates $B^1$ with the its averaged part:
\[
B=[B^1]= \left(\begin{array}{cc}
1 & \; [b^1_{12}] \\
0 & \; 1  
\end{array}\right)
\]
which is in the form of (\ref{eq:floquetform}).\qed

Thus, applying this lemma, the flow (\ref{eq:skewdebil}) is reducible to
constant coefficients with Floquet matrix $B$, of the form
(\ref{eq:floquetform}). That is, there exists a real analytic 
map $Z: \mT \to SL(2,\mR)$ such that
\begin{equation}\label{eq:conjugationdebil}
A(\theta) Z(\theta) = Z(\theta+2\pi\omega) B
\end{equation}
for all $\theta \in \mT$. Moreover, since the trace of $B$ 
is $2$, the rotation
number of (\ref{eq:partdual}) is rational, so that we are at
the endpoint of a gap, which we want to show that is
non-collapsed.

By the arguments of Section \ref{sec:prerequisites}, we rule out 
the possibility of $B$ being the identity. Indeed, this
would imply the coexistence of two quasi-periodic analytic solutions
with frequency $\omega$, which does happen in the Almost
Mathieu case. Therefore $B\ne I$ and,
thus, $c\ne 0$ in the definition above.

If $B \ne I$, it is a well-known fact of Floquet theory that $2 a^k/b$
lies at the endpoint of a non-collapsed gap (see, for example, the monograph 
\cite{yaku-star} for classical Floquet theory or \cite{broer-puig-simo} 
for the continuous and quasi-periodic Schr\"odinger case). For the sake of 
self-completeness we sketch the argument.

We will see that there exists a $\alpha_0>0$ such that if
$0<|\alpha|<\alpha_0$ and $\alpha$ is either positive or negative
(depending on the sign of $c$) then $2 a^k/b + \alpha$ lies in the
resolvent set of $\sigma_{4/b}$. To do so, we will show that, for
these values of $\alpha$, the skew-product flow
\begin{equation}\label{eq:skewperturb}
\left(\begin{array}{c}
x_{n+1} \\
x_{n} 
\end{array}\right)=
\left(
\begin{array}{cc}
\frac{2 a^k}{b} +\alpha - \frac{4}{b} \cos{\theta_n}  & \;-1 \\
1 & \;\fm 0
\end{array}
\right)
\left(\begin{array}{c}
x_{n} \\
x_{n-1} 
\end{array}\right),\qquad \theta_{n+1} = \theta_n + 2\pi \omega
\end{equation}
has an exponential dichotomy (see Coppel \cite{coppel}) which implies
that $2 a^k/b + \alpha \not\in \sigma_{4/b}$ (see Johnson
\cite{johnson-82}). The reduction given by $Z$ transforms this system
into
\begin{eqnarray}\label{eq:rotating}
y_{n+1} & = &
\left(
\begin{array}{cc}
1 + \alpha\left( z_{11} z_{12} - c z_{11}^2 \right)
	& \;\;\;  c + \alpha\left(-c z_{11} z_{12} + z_{12}^2\right) \\
-\alpha z_{11}^2  & \;\;\; 1 - \alpha z_{11} z_{12}
\end{array}
\right)
y_n, \nonumber \\
\theta_{n+1} & = & \theta_n + 2\pi \omega,
\end{eqnarray}
where $y_n \in \mR^2$ are the new variables. The $z_{ij}$ are
the elements of the matrix $Z$ and we have used the relations given
by (\ref{eq:conjugationdebil}) and the special form of $A$
and $B$. In the same calculation, we also see that 
$\left(z_{11}(2\pi n\omega)\right)_{n\in \mZ}$ is a
quasi-periodic solution of equation (\ref{eq:partdual}) and that
it is not identically zero. Using averaging theory (see, for
example, \cite{arnold:geometrical}), system
(\ref{eq:rotating}) can be transformed into 
\begin{eqnarray}\label{eq:averaged}
y_{n+1} & = &
\left(\left(
\begin{array}{cc}
1 + \alpha\left( [z_{11} z_{12}] - c [z_{11}^2] \right)
	&  \;\;\; c + \alpha\left(-c [z_{11} z_{12}] + [z_{12}^2]\right) \\
-\alpha [z_{11}^2]  & \;\;\; 1 - \alpha [z_{11} z_{12}]
\end{array}
\right)+ M\right)
y_n \nonumber \\
\theta_{n+1} & = & \theta_n + 2\pi \omega
\end{eqnarray}
by means of a conjugation in $SL(2,\mR)$, with $M$ analytic in both
$\theta$ and $\alpha$ (in some narrower domains) and of order $\alpha^2$. The
time-independent part of the above system is hyperbolic if
$c\alpha<0$. Therefore, if $|\alpha|\ne 0$ is small enough the
time-dependent system (\ref{eq:averaged}) has an exponential
dichotomy for $c\alpha<0$. Hence $2a^k/b + \alpha$ does not belong to
$\sigma_{4/b}$. Since this works for all $a^k$, (which are dense in the
spectrum), $\sigma_{4/b}$ is a Cantor set. By duality the result is also
true for $\sigma_{b}$. This ends the proof of Corollary
\ref{res:debil}. 

\begin{remark}
The same can be done for the operator $H_{b,\pi/2}$ instead
of $H_{b,0}$. In this case the Floquet matrix has trace
$-2$. The corresponding point eigenvalues correspond to ends
of non-collapsed gaps and are dense in the spectrum.
\end{remark}

\section*{Acknowledgements}
The author is indebted to Hakan Eliasson, Raphael Krikorian
and Carles Sim\'o for stimulating discussions and comments
on this problem. He would like to thank the anonymous
referee for his useful suggestions. He also wants to thank  the 
\emph{Centre de Math\'ematiques} at the
\emph{\'Ecole Polytechnique} for hospitality. Help from the Catalan
grant 2000FI71UBPG and grants DGICYT BFM2000-805 
(Spain) and CIRIT 2000 SGR-27, 2001 SGR-70 (Catalonia) is also
acknowledged.

\def\cprime{$'$}

\end{document}